\documentclass[preprint,prl,showpacs,floatfix]{revtex4}

\usepackage{graphicx}
\usepackage{dcolumn}
\usepackage{bm}
\usepackage{amsmath}

\begin{document}

\title{%
Stabilized-jellium description of neutral and multiply charged
fullerenes C$^{x\pm}_{60}$} 
\author{C. Yannouleas and Uzi Landman}
\affiliation{%
School of Physics, Georgia Institute of Technology,
Atlanta, Georgia 30332-0430 }
\date{August 1993; Chem. Phys. Lett. {\bf 217}, 175 (1994)}

\begin{abstract}
A description of neutral and multiply charged 
fullerenes is proposed based on a stabilized jellium
(structureless pseudopotential) 
approximation for the ionic background and the local density
approximation for the $\sigma$ and $\pi$ valence electrons.
A recently developed shell-correction method is used to
calculate total energies and properties of both the neutral
and multiply charged anionic and cationic fullerenes.
The effect of the icosahedral symmetry is included perturbatively. 
The calculated single-particle energy level spectrum of C$_{60}$ is
in good correspondence with experimentally measured ones and previous
self-consistent local-density-approximation calculations.
For the multiply charged fullerenes, we 
calculate microscopically the charging energies of C$_{60}^{x \pm}$ for up to
$x=12$ excess charges. A semiclassical interpretation of these results
is developed, which views the fullerenes as Coulomb islands possessing
a classical capacitance. The calculated values for the first ionization 
potential and the first electron affinity agree well with the experimental 
ones. For the second and third ionization potentials, there exist substantial 
discrepancies in the experimental measurements. 
Our calculations support the results from charge transfer bracketing 
experiments and from direct ionization experiments through electron impact.
The doubly charged negative ion is found to be a very long-lived metastable 
species, in agreement with observations.
\end{abstract}

\maketitle

\newpage

\section{introduction}

Charging of macroscopic metal spheres is an old subject with scientific
accounts dating back to Coulomb, Faraday, and others \cite{fara}. 
Recently, issues related to electrical charging emerged in
connection with quantal nanostructures which are of interest 
to diverse areas of condensed-matter and atomic, molecular and cluster
physics, such as nanofabricated semiconductor devices known as 
Coulomb islands \cite{kast1,kast2}, metal microclusters 
\cite{yl1,yl2,brec,sneg}, electron attachment to molecular
\cite{coe,barn} and alkali-halide \cite{scha} clusters,
and carbon clusters and fullerenes 
\cite{elva,baba,berk,walt,steg,hett,limb,schau}.

We report on a unified theoretical study of the energetics, stability,
ionization potentials and electron affinities of neutral and
charged carbon fullerenes, C$^{x\pm}_{60}$, using local density
functional theory (LDA) \cite{ks}
with a {\it stabilized\/}-jellium background \cite{perd}.
The present approach is an adaptation to the case of fullerenes of a
shell-correction method (SCM), developed by us previously for metal 
microclusters
\cite{yl1,yl2}. The main elements of this adaptation consist of
employment of the stabilized jellium instead of the usual jellium
background, and the use of a generalized electron-density profile
(appropriate for description of the fullerene cage).
Furthermore, the point-group icosahedral symmetry of the C$_{60}$
cage is introduced via a perturbative treatment.

The electronic structure of fullerenes has been investigated extensively
using various methods ranging from {\it ab initio\/} all-electron 
quantum chemical Hartree-Fock calculations, and self-consistent Kohn-Sham LDA
calculations in conjunction with nonlocal pseudopotentials or a
jellium background for the ions, to simplified free-electron \cite{kuhn,gall}
and particle-on-a-sphere models \cite{savi}. Our generalized shell-correction
method (GSCM) which combines elements of the LDA methodology with
simplifications circumventing self-consistent solution of the Kohn-Sham
(KS) equations represents a substantial improvement over certain previous
simplified methods, yielding results in quantitative agreement with 
self-consistent KS-LDA {\it ab initio} pseudopotential 
calculations (see, e.g., \cite{trou,rose,ye}), and experimental data.

Additionally, our method permits us to investigate, 
within the local density approximation, the important
class of multiply anionic fullerene systems, where the familiar KS-LDA
is known to fail \cite{yl1,yl2}.
Until now, in all cases but one 
\cite{pede}, properties of anionic fullerenes (whose energetics is relevant
not only for
understanding the properties of free molecules, but also those of
fullerite intercalation compounds \cite{wang} and complexes \cite{cios})
have been treated within the Hartree-Fock approximation 
\cite{hett,cios,hutt} or estimated with a simplified classical electrostatic
model \cite{gall,wang}, associated with the particle-on-a-sphere model.
Both these methods, however, omit essential theoretical ingredients,
namely the correlation energy in the case of the former, and the shell 
structure and appropriate radial and angular density distribution of the
electronic charge in the case of the latter.
The recent study by Pederson and Quong \cite{pede},
using an LDA all-electron full-potential Gaussian-orbital basis, was
performed within the ansatz of a localized basis expansion of the 
Kohn-Sham orbitals, yielding results in good agreement with those 
obtained by us using the GSCM.

\section{LDA theoretical method}

\subsection{Stabilized jellium approximation}

Fullerenes and related carbon structures
have been extensively investigated using {\it ab initio\/}
local-density-functional
methods and self-consistent solutions of the Kohn-Sham (KS)
equations \cite{trou,koha}.
For metal clusters, replacing
the ionic cores with a uniform jellium background was found to
describe well their properties within the KS-LDA method
(see references in Refs. \cite{yl1,yl2}). 
Motivated by these results, several attempts
to apply the jellium model in conjunction with LDA to investigations
of fullerenes have appeared recently \cite{yaba,lipp,pusk}.
Our approach differs from the earlier ones in several aspects and,
in particular, in the adaptation to the case of finite systems 
of the stabilized-jellium (or structureless pseudopotential)
energy density functional (see eq.\ (\ref{epseu}) below and
Ref.\ \cite{perd}).\footnote{
Other aspects differentiating our study from previous ones
are: Consideration of the icosahedral perturbation,
unlike Refs.\ \cite{lipp,pusk}; consideration of 240 active electrons
(as is the case with {\it ab initio\/} 
LDA pseudopotential  claculations, see i.e.,
Ref.\ \cite{trou}), unlike Ref.\ \cite{yaba} which considered 360 active 
electrons, Ref.\ \cite{lipp} which considered 60 active electrons,
and Ref.\ \cite{pusk} which considered 250 active electrons.}

An important shortcoming of the standard jellium approximation for
fullerenes (and other systems with high density , i.e., small
$r_s$) results from a well-known property of the jellium at
high electronic densities, namely that the jellium is unstable and 
yields negative surface-energy contribution to the total energy
\cite{perd},
as well as unreliable values for the total energy.
These inadequacies of the standard jellium model can be rectified by
pseudopotential corrections. 
A modified-jellium approach
which incorporates such pseudopotential corrections
and is particularly suited for our purposes here, is the 
{\it structureless pseudopotential\/} model or
{\it stabilized jellium\/} approximation developed in Ref.\ \cite{perd}.

In the stabilized jellium, the total energy $E_{pseudo}$, as a
functional of the electron density $\rho({\bf r})$, is given
by the expression
\begin{equation}
E_{pseudo}[\rho,\rho_+] = E_{jell}[\rho,\rho_+] +
\langle \delta \upsilon \rangle_{WS}
\int \rho({\bf r}) {\cal U}({\bf r}) d{\bf r}
-\widetilde{\varepsilon} \int \rho_+({\bf r}) d{\bf r}~,
\label{epseu}
\end{equation}
where by definition the function ${\cal U}({\bf r})$ equals unity inside, 
but vanishes, outside the jellium volume. 
$\rho_+$ is the density of the positive jellium background
(which for the case of C$_{60}$ is taken as a spherical shell,
of a certain width $2d$, centered at 6.7 $a.u.$ ).
$E_{pseudo}$ in eq.\ (\ref{epseu}) is the standard jellium-model total energy, 
$E_{jell}$, modified by two corrections. The first correction adds the effect
of an average (i.e., averaged over the volume of a Wigner-Seitz cell) 
difference potential, $\langle \delta \upsilon \rangle_{WS} \cal{U} ({\bf r})$, 
which acts on the electrons in addition to the standard jellium attraction
and is due to the atomic pseudopotentials (in this work, we use the
Ashcroft empty-core pseudopotential, specified by a core radius
$r_c$, as in Ref.\ \cite{perd}).
The second correction subtracts 
from the jellium energy functional the spurious electrostatic
self-repulsion of the positive background within each cell; this term makes
no contribution to the effective electronic potential.

Following Ref.\ \cite{perd}, the bulk stability condition (eq.\ (25) in
Ref.\ \cite{perd}) determines the value of the pseudopotential core
radius $r_c$, as a function of the bulk Wigner-Seitz radius $r_s$.
Consequently, the difference potential can be expressed
solely as a function of $r_s$ as follows (energies in $Ry$, 
distances in $a.u.$):
\begin{equation}
\langle \delta \upsilon \rangle_{WS} = 
-\frac{2}{5} \left ( \frac{9\pi}{4}\right )^{2/3} r_s^{-2} + 
\frac{1}{2\pi} \left ( \frac{9\pi}{4} \right )^{1/3} r_s^{-1} +
\frac{1}{3} r_s \frac{d\varepsilon_c}{dr_s}~,
\label{dvws}
\end{equation}
where
$\varepsilon_c$ is the per particle electron-gas correlation energy
(in our calculation, we use the Gunnarsson-Lundqvist 
exchange and correlation energy functionals (see Refs.\ \cite{yl1,yl2})).

The electrostatic self-energy, $\widetilde{\varepsilon}$, per unit charge 
of the uniform positive jellium is given by
\begin{equation}
\widetilde{\varepsilon} = 6\upsilon^{2/3}/5r_s~,
\label{vare}
\end{equation}
where $\upsilon$ is the valence of the atoms ($\upsilon=4$ for carbon).

\subsection{Shell-correction method}

Besides the familiar KS-LDA approach,
an alternative LDA method, which has been used in studies
of metal clusters \cite{yl1,yl2,yann},
is based on an
extended Thomas Fermi (ETF) variational
procedure using a parametrized density profile $\rho({\bf r}; \{\gamma_i\})$,
with $\{\gamma_i\}$ as variational parameters \cite{yl2}.
In the case of metal clusters, the energy density functional that 
is variationally minimized consists of a kinetic-energy functional 
$T[\rho]$, including terms up to 4th order in the density
gradients (see Ref.\ \cite{yl2}),
and of {\it potential\/} terms according to the 
standard jellium-LDA functional. The effective single-particle 
potentials and associated single-particle energy spectra 
obtained by this method for metal clusters
provide a good approximation to the corresponding ones obtained 
from KS-LDA calculations \cite{yl1,yl2}, 
and have been extensively used \cite{yann} in studies  
of the optical properties of metal clusters.
To apply the ETF-LDA method to carbon fullerenes, we generalize it by 
employing potential terms according to the
stabilized-jellium functional (\ref{epseu}).

Another required generalization consists in employing
a parametrized electron-density profile that accounts for the hollow
cage-structure of the fullerenes. 
Such a density profile is provided by the following adaptation of a 
generalization of an inverse Thomas-Fermi distribution,
used earlier in the context of nuclear physics \cite{gram}, i.e.,
\begin{equation}
\rho (r) = \rho_0 
\left( 
\frac{ F_{i,o} \sinh [w_{i,o}/\alpha_{i,o}]}
{\cosh [w_{i,o}/\alpha_{i,o}] + \cosh [(r-R)/\alpha_{i,o}]} 
\right)^{\gamma_{i,o}}~,
\label{rgram}
\end{equation}
where $R=6.7\;a.u.$ is the radius of the fullerene cage.
$w$, $\alpha$, and $\gamma$ are variables to be determined
by the ETF-LDA minimization. For $R=0$ and large values of $w/\alpha$,
expression (\ref{rgram}) approaches the more familiar 
inverse Thomas-Fermi distribution, with
$w$ the width, $\alpha$ the diffuseness and $\gamma$
the asymmetry of the profile around $r=w$.
There are a total of six parameters to be determined, since
the indices $(i,o)$ stand for the regions inside ($r < R$) and 
outside ($r>R$) the fullerene cage. 
$F_{i,o} = (\cosh [w_{i,o}/\alpha_{i,o}]+1)/\sinh [w_{i,o}/\alpha_{i,o}]$ 
is a constant guaranteeing that the two parts of the curve join smoothly
at $r = R$.
The density (\ref{rgram}) peaks at $r=R$ and then falls towards
smaller values both inside and outside the cage 
(see top panel of Fig.\ 1).

\begin{figure}[t]
\centering\includegraphics[width=6.5cm]{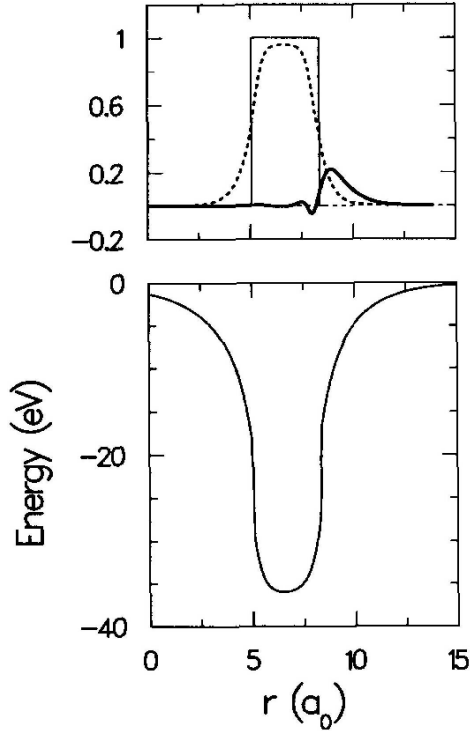}
\caption{
Bottom panel: The stabilized-jellium LDA potential obtained by the ETF
method for the neutral C$_{60}$ molecule. The Wigner-Seitz radius for
the jellium bacground is 1.23 $a.u.$ Note the asymmetry of the potential
about the minimum. The associated difference potential
$\protect \langle \delta \upsilon \protect\rangle_{WS}=-9.61 \; eV$.
\protect\\
Top panel: Solid line: Radial density of the positive jellium background.
Dashed line: ETF electronic density. Note its asymmetry about the maximum.
Thick solid line: The difference (multiplied by 10) of electronic ETF densities
between C$_{60}^{5-}$ and C$_{60}$. It illustrates that the excess
charge accumulates in the outer perimeter of the total electronic
density. All densities are normalized to the density of the positive
jellium background.
}

\end{figure}

However, apart from the effective potentials, 
other properties of metal clusters -- and by analogy of
fullerenes -- determined via the ETF-LDA 
method, such as ionization potentials (IPs), electron affinities
(EAs), and total energies, compare with the 
self-consistently calculated KS-LDA values only in an average
sense, i.e., they do not exhibit shell effects. 
In our present study, shell effects are incorporated in the GSCM according 
to the procedure of Refs.\ \cite{yl1,yl2}. As elaborated in Ref.\ \cite{yl2},
shell effects in the total energy are contained, 
to first order in
$\delta \rho_{\text{KS}}$  
($\delta \rho_{\text{KS}} = \rho_{\text{KS}}-\widetilde{\rho}$, where
$\widetilde{\rho}$ is the ETF-LDA optimized density specified above
and $\rho_{\text{KS}}$ is the self-consistent KS density),
in the sum $\sum_i \widetilde{\varepsilon_i}$, 
where $\widetilde{\varepsilon_i}$ are the single-particle energies of the
ETF effective potential, and the sum extends over the occupied levels.
Consequently, we replace the kinetic-energy 
$\widetilde{T}$ in the ETF-LDA functional by
\begin{equation}
T_{\text{sh}} = \sum_i \widetilde{\varepsilon}_i 
-\int \widetilde{\rho} ({\bf r}) 
\widetilde{V}({\bf r}; \widetilde{\rho}({\bf r})) d{\bf r}~,
\label{tsh}
\end{equation}
where $\widetilde{V}({\bf r}; \widetilde{\rho}({\bf r}))$ is the effective
potential produced by the ETF method \cite{yl1,yl2} (as an example,
see lower panel Fig.\ 1 for the ETF potential 
associated with the neutral C$_{60}$).
As a result, the total energy, $E_{\text{sh}}$, including the shell
correction, $\Delta E_{\text{sh}}=T_{\text{sh}}-\widetilde{T}$, is given by 
$E_{\text{sh}}[\widetilde{\rho}]=
T_{\text{sh}} - \widetilde{T} + \widetilde{E}[\widetilde{\rho}]$,
where $\widetilde{E}$ is the ETF-LDA energy-density functional.

After some rearrangenments, the shell-corrected total energy 
$E_{\text{sh}}[\widetilde{\rho}]$ in the GSCM can be written in functional 
form as follows
\begin{equation}
E_{\text{sh}}[\widetilde{\rho}]=
\sum_i \widetilde{\varepsilon}_i
- \int \! \left\{ 
\frac{1}{2} \widetilde{V}_H({\bf r}) + \widetilde{V}_{\text{xc}}({\bf r})
\right\}
\widetilde{\rho}({\bf r}) d\/{\bf r} 
+ \int \widetilde{{\cal E}}_{\text{xc}} 
[\widetilde{\rho}({\bf r})] d\/{\bf r}
+ E_I - \widetilde{\varepsilon} \int \rho_+({\bf r}) d{\bf r}~,
\label{ensh}
\end{equation}
where $\widetilde{V}_H$ and $\widetilde{V}_{\text{xc}}$ are the Hartree and 
exchange-correlation  electronic potentials, $\widetilde{{\cal E}}_{\text{xc}}$ 
is the exchange-correlation energy density functional, 
and $E_I$ is the energy of the positive jellium background.
The specific way of writing the functional $E_{\text{sh}}$ above was chosen 
so that its similarity in form to the Harris functional \cite{harr}
is evident. We note that our method differs from that approach in that
the optimization of the input density is achieved by us through a variational
ETF method, which does not require a step-by-step matrix diagonalization.
While our focus in the previous papers \cite{yl1,yl2} was on 
jellium models for metal clusters,
the very good agreement between our results and those obtained via 
self-consistent KS-LDA jellium calculations \cite{yl1,yl2}
suggested that it 
would be worthwhile to explore the application of our SCM to more general 
electronic systems extending beyond metal clusters.
The present study is an example of such an application of the SCM to
fullerenes.

\subsection{Icosahedral splitting}

Heretofore, the point-group icosahedral symmetry of C$_{60}$
was not considered, since the molecule was treated as a spherically symmetric 
cage. This is a reasonable zeroth-order approximation as noticed by several
authors \cite{trou,gall,pusk,hadd}. However, considerable improvement
is achieved when the effects of the point-group icosahedral symmetry 
are considered as a next-order correction (mainly the
lifting of the angular momentum degeneracies).

The method of introducing the icosahedral splittings is that of the
crystal field theory. Thus, we will use the fact that the bare electrostatic
potential from the ionic cores, considered as point charges, 
acting upon an electron, obeys the
well-known expansion theorem \cite{gerl}
\begin{equation}
U({\bf r}) = - \upsilon e^2 \sum_i \frac{1}{|{\bf r}- {\bf r}_i|} =
- \sum_{l=0}^{\infty} \sum_{m=-l}^{l} \kappa_l (r) C_l^m Y_l^m (\theta,\phi)~,
\label{cryf}
\end{equation}
where the angular coefficients $C_l^m$ are given through the angular
coordinates $\theta_i, \phi_i$ of the carbon atomic cores, namely,
\begin{equation}
C_l^m = \sum_i Y_l^{m*} (\theta_i, \phi_i)~,
\label{clm}
\end{equation}
where $*$ denotes complex conjugation.

\begin{figure}[t]
\centering\includegraphics[width=6.5cm]{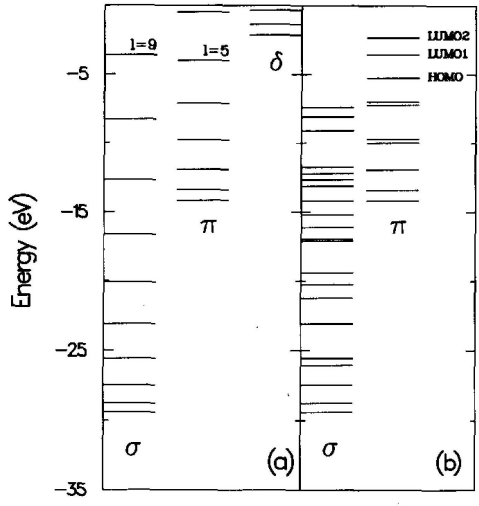}
\caption{
(a) The single-particle levels of the ETF-LDA potential for C$_{60}$ 
shown in Fig. 1. Because of the spherical symmetry, they are characterized
by the two principle quantum numbers $n_r$ and $l$, where $n_r$ is the
number of radial nodes and $l$ the angular momentum. They are grouped 
in three bands labeled $\sigma$ ($n_r=0$), $\pi$ ($n_r=1$), and $\delta$ 
($n_r=2$). Each band starts with an $l=0$ level. \protect\\
(b) The single-particle levels for C$_{60}$ after the icosahedral
splittings are added to the spectra of (a). The tenfold degenerate 
HOMO (h$_u$) and the sixfold degenerate LUMO1 (t$_{1u}$) and LUMO2 
(t$_{1g}$) are denoted; they originate from the spherical $l=5$ 
and $l=6$ (t$_{1g}$) $\pi$ levels displayed in panel (a).
For the $\sigma$ electrons, the icosahedral perturbation strongly splits
the $l=9$ level of panel (a). There result five sublevels which straddle the
$\sigma$-electron gap as follows: two of them (the eightfold degenerate g$_u$, 
and the tenfold degenerate h$_u$) move down and are fully occupied
resulting in a shell closure (180 $\sigma$ electrons in total). 
The remaining unoccupied levels, originating 
from the $l=9$ $\sigma$ level, are sharply shifted upwards and
acquire positive values.
}
\end{figure}

We take the radial parameters $\kappa_l(r)$ as constants, and determine
their value by adjusting the icosahedral single-particle spectra 
$\varepsilon_i^{\text{ico}}$ to reproduce 
the pseudopotential calculation of Ref.\ \cite{trou},
which are in good agreement with experimental data.
Our spectra without and with icosahedral splitting are shown in Fig.\ 2a
and Fig.\ 2b, respectively.
We find that a close reproduction of the results of {\it ab initio\/}
LDA calculations \cite{trou,rose,ye} is achieved when 
the Wigner-Seitz radius for the jellium background is $1.23$ $a.u.$
The shell corrections including the icosahedral splittings are 
calculated using the icosahedral single-particle energies 
$\varepsilon_i^{\text{ico}}$ in eq.\ (\ref{tsh}). The average quantities
($\widetilde{\rho}$ and $\widetilde{V}$) are
maintained as those specified through the ETF variation
with the spherically symmetric profile of eq.\ (\ref{rgram}).
This is because the first-order correction to the total energy
(resulting from the icosahedral
perturbation) vanishes, since the integral over the sphere of a spherical
harmonic $Y_{l}^{m}\;(l>0)$ vanishes.

\section{Results}

\subsection{Ionization potentials and electron affinities}

Having specified the appropriate Wigner-Seitz radius $r_s$ and 
the parameters $\kappa_l$ 
of the icosahedral crystal field through a comparison with the 
pseudopotential LDA calculations for the neutral C$_{60}$, we calculate 
the total energies of the cationic and anionic species by allowing for a 
change in the total electronic charge, namely by imposing the constraint
\begin{equation}
4 \pi \int \rho(r) r^2 dr =240 \pm x~,
\label{norm}
\end{equation}
where $\rho(r)$ is given by eq.\ (\ref{rgram}).
The shell-corrected and icosahedrally perturbed
first and higher ionization potentials $I_x^{\text{ico}}$
are defined as the difference of the ground-state
shell-corrected total energies $E_{\text{sh}}^{\text{ico}}$ as follows:
\begin{equation}
I_x^{\text{ico}} = E_{\text{sh}}^{\text{ico}}(N_e=240-x; Z=240)
-E_{\text{sh}}^{\text{ico}}(N_e=240-x+1; Z=240)~,
\label{ips}
\end{equation}
where $N_e$ is the number of electrons in the system and $x$ is the number 
of excess charges on the fullerenes (for the excess charge, we will 
find convenient to use two different notations $x$ and $z$ related as
$x=|z|$. A negative value of $z$ corresponds to positive excess charges).
$Z=240$ denotes the total positive charge of the jellium background.

The shell-corrected and icosahedrally perturbed first 
and higher electron affinities $A_x^{\text{ico}}$ are similarly defined as 
\begin{equation}
A_x^{\text{ico}} = E_{\text{sh}}^{\text{ico}}(N_e=240+x-1; Z=240)
-E_{\text{sh}}^{\text{ico}}(N_e=240+x; Z=240)~.
\label{eas}
\end{equation}

We have also calculated the corresponding average quantities
$\widetilde{I}_x$ and $\widetilde{A}_x$, which result from the ETF
variation with spherical symmetry (that is without shell and icosahedral
symmetry corrections). Their definition is the same as in
Eqs.\ (\ref{ips}) and (\ref{eas}), but with the index $sh$ replaced by
a tilde and the removal of the index $ico$. 

In our calculations of the charged fullerene molecule, the $r_s$ value
and the icosahedral splitting parameters ($\kappa_l$, see Eq. (\ref{cryf}),
and discussion below it) were taken as those which were determined
by our calculations of the neutral molecule, discussed in the previous
section. The parameters which specify the ETF electronic density 
(Eq. (\ref{rgram})) are optimized for the charged molecule,
thus allowing for relaxation effects due to the excess charge. This
procedure is motivated by results of previous electronic structure
calculations for C$_{60}^+$ and C$_{60}^-$ \cite{rose,ye}, 
which showed that the icosahedral 
spectrum of the neutral C$_{60}$ shifts almost rigidly upon charging
of the molecule.

\begin{table}
\caption{
ETF (spherically averaged, denoted by a tilde) and shell-corrected 
(denoted by a superscript $ico$ to indicate that the icosahedral
splittings of energy levels have been included).
IPs and EAs of fullerenes C$^{x \pm}_{60}$. Energies in $eV$. 
$r_s=1.23$ $a.u.$}
\begin{tabular}{ccccc}
$x$ & $\widetilde{I}_x$ & $I^{\text{ico}}_x$ &
$\widetilde{A}_x$ & $ A^{\text{ico}}_x$ \\ \hline
1 &  5.00  &  7.40 & 2.05 & 2.75 \\
2 &  7.98  & 10.31 & $-$0.86 & $-$0.09 \\
3 & 10.99  & 13.28 & $-$3.75 & $-$2.92 \\
4 & 14.03  & 16.25 & $-$6.60 & $-$5.70 \\
5 & 17.09  & 19.22 & $-$9.41 & $-$8.41 \\
6 & 20.18  & 22.20 & $-$12.19 & $-$11.06 \\
7 & 23.29  & 25.24 & $-$14.94 & $-$14.85 \\
8 & 26.42  & 28.31 & $-$17.64 & $-$17.24 \\
9 & 29.57  & 31.30 & $-$20.31 & $-$19.49 \\
10 & 32.73 & 34.39 & $-$22.94 & $-$21.39 \\
11 & 35.92 & 39.36 & $-$25.53 & $-$22.93 \\
12 & 39.12 & 42.51 & $-$28.07 & $-$23.85 
\end{tabular}
\end{table}

Shell-corrected and ETF calculated values of ionization potentials (IPs)
and electron affinities (EAs), for values of the
excess charge up to 12 units, are summarized in Table I (for $r_s=1.23\; a.u.$).
Our shell-corrected results for the first electron affinity and first 
ionization potential are in good agreement with the experimental values
(2.75 $eV$ \cite{yang} and 7.54 $eV$ \cite{hert}, respectively). 
For the second affinity, we find a small negative value of $-0.09$ $eV$
indicating that C$^{2-}_{60}$ is a metastable species (for lifetime
estimates, see below). For the second ionization potential our value
of 10.31 $eV$ is close to that determined experimentally by McElvany 
{\it et al.\/} (9.7 $eV$) \cite{elva} using charge-transfer bracketing
experiments, and most recently by Sai Baba {\it et al.\/} (10.3 $eV$) 
\cite{baba} using electron impact ionization. There is a controversy about 
this value,
since several other experiments have found values that vary from 8.5 $eV$
to 12.25 $eV$ (for a detailed review of related experimental measurements
available in the literature, cf. Ref.\ \cite{baba}, and references
therein). Our theoretical calculations support the 
measurements of McElvany {\it et al.\/} and Sai Baba {\it et al.\/},
epecially since another quantity -- namely, the appearance energy of 
C$_{60}^{3+}$, that is, the sum of $\sum_{x=1}^3 I^{\text{ico}}_x$ -- 
which was measured in the recent work of Ref.\ \cite{baba} 
is in very good agreement with the value calculated by us.
Indeed, the appearance energy of C$_{60}^{3+}$ determined in 
Ref.\ \cite{baba} by linear extrapolation of the ionization efficiency curves 
was found to be 33.2$\pm$1 $eV$. This value should be considered as the upper 
limit, due to a curvature at the foot of the ionization efficiency curve. Ref.\
\cite{baba} estimates that this curvature leads to an overestimation of
2 $eV$. Taking this correction into account, the resulting experimental
value of 31.2$\pm$1 $eV$ is in close agreement with our theoretical value
of 30.99 $eV$.

\begin{figure}[t]
\centering\includegraphics[width=6.5cm]{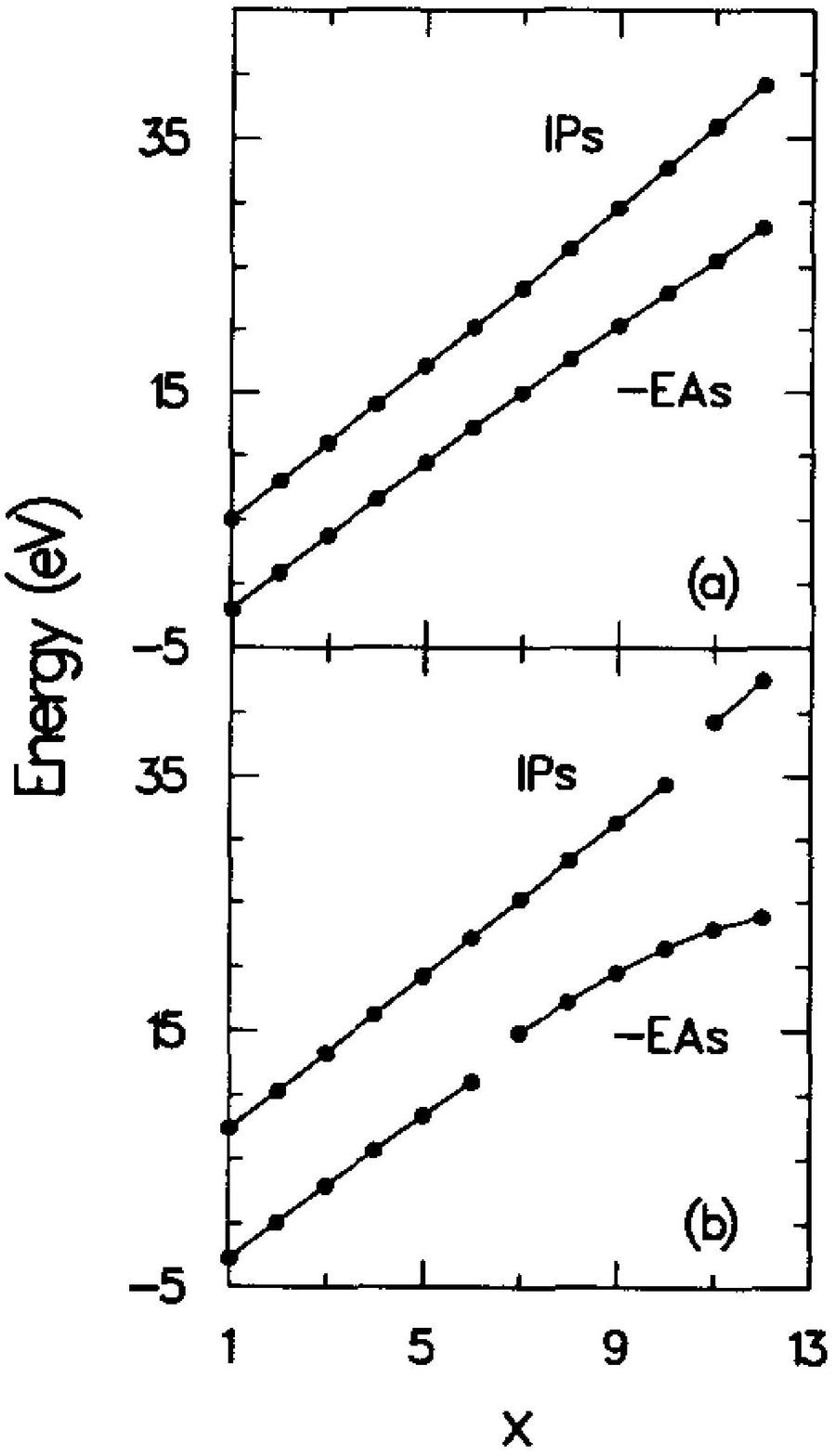}
\caption{
(a) Extended Thomas-Fermi LDA Ionization potentials ($\widetilde{I}_x$) and 
electron affinities ($-\widetilde{A}_x$)
as a function of the excess charge $x$. \protect \\
(b) Shell-corrected icosahedral IPs ($I^{\text{ico}}_x$) and 
$-$EAs ($-A^{\text{ico}}_x$) as a function of the excess charge $x$.
}
\end{figure}

The IPs and EAs calculated by us are plotted against the excess charge $x$
in Fig.\ 3.  For the ETF results (see Fig.\ 3a) the 
dependence is linear to a remarkable degree. Small deviations from
linearity, however, are discernible and are due to the
variations of electronic spill-out with varying excess charge.
An inspection of the shell-corrected results in Fig.\ 3b reveals
that they also exhibit a linear relationship with $x$ within each
electronic shell. Discontinuities due to shell openings are clearly
discernible [ $1 \leq x \leq 10$ for IPs corresponds to removal of
electrons from the same h$_u$ shell, while $11 \leq x \leq 12$ 
corresponds to removal of an electron from the shell immediately below it 
(g$_g$ shell); $1 \leq x \leq 6$ for the EAs corresponds
to adding an electron to the same t$_{1u}$ shell, while $7 \leq x \leq 12$
corresponds to an extra electron in the same t$_{1g}$ shell. The gap
between the t$_{1u}$ and the t$_{1g}$ shell is approximately 1.23 $eV$,
while the gap between the h$_u$ shell and the 
g$_g$ shell is 1.75 $eV$].  The bending of the
shell-corrected affinities around $x=9-12$ is due to the fact that
the electron emission for such highly charged states of the free molecule 
is barrierless and thus the uncertainty of the LUMO 
energy is high (see below). In the case of intercalated or endohedral 
compounds, the LUMO is drastically more bounded,
due to the attraction of the ionic intercalant, and one should expect
a strong linear relation even for this range of excess charge.

\begin{figure}[t]
\centering\includegraphics[width=6.5cm]{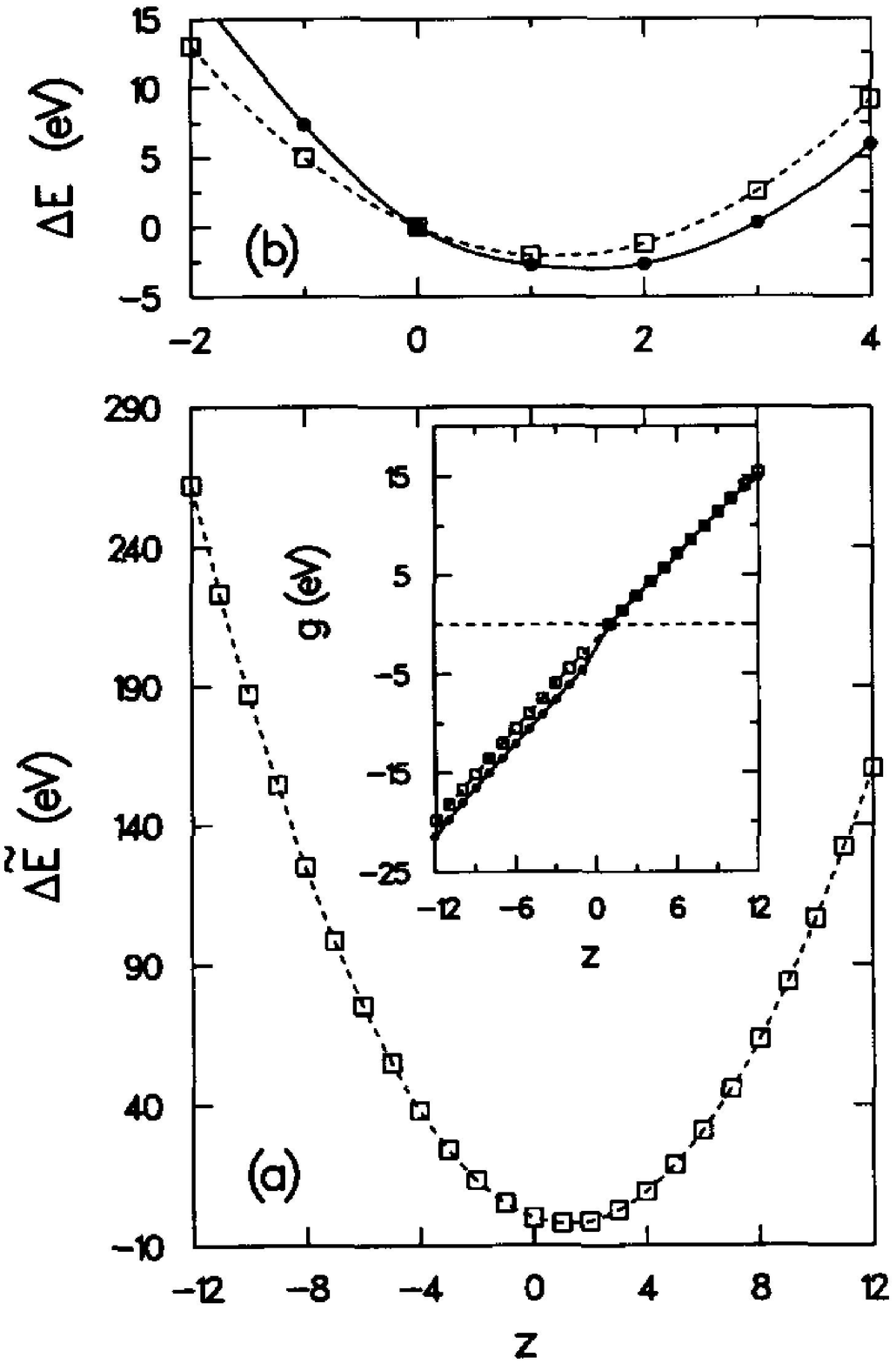}
\caption{
(a) ETF-LDA total energy differences (appearance energies)
$\Delta \widetilde{E}(z)=\widetilde{E}(z)-\widetilde{E}(0)$
as a function of the excess charge $z$ 
($z<0$ corresponds to positive excess charge). \protect \\
Inset: The ETF function $\widetilde{g}(z)$ (open squares),
and the shell-corrected 
function $g^{\text{ico}}_{\text{sh}}(z)$ (filled circles).
For $z \geq 1$ the two functions are almost identical. \protect \\
(b) magnification of the appearance-energy curves
for the region $-2 \leq z \leq 4$. Filled circles: shell-corrected
icosahedral values 
[$\Delta E^{\text{ico}}_{\text{sh}}(z)
=E^{\text{ico}}_{\text{sh}}(z)-E^{\text{ico}}_{\text{sh}}(0)$].
Open squares: ETF-LDA values
[$\Delta \widetilde{E}(z)=\widetilde{E}(z)-\widetilde{E}(0)$].
}
\end{figure}

\subsection{Charging energies and capacitance}

Fig.\ 4a shows that the variation of the total ETF-LDA energy 
difference (appearance energies)
$\Delta \widetilde{E}(z)=\widetilde{E}(z)-\widetilde{E}(0)$, as a 
function of excess charge $z$ ($|z|=x$), exhibits a parabolic behavior.
The inset in Fig.\ 4a exhibiting the quantity 
\begin{equation}
\widetilde{g}(z) = \frac{\widetilde{E}(z)-\widetilde{E}(0)}{z}
+ \widetilde{A}_1~,
\label{gsm}
\end{equation}
plotted versus $z$ (open squares), shows 
a straight line which crosses the zero energy line at $z=1$. As a result
the total ETF-LDA energy has the form,
\begin{equation}
\widetilde{E}(z) = \widetilde{E}(0) + \frac{z(z-1)e^2}{2C}
- \widetilde{A}_1 z~.
\label{etfe}
\end{equation}
Equation (\ref{etfe}) indicates that fullerenes behave on the average like
a capacitor having a capacitance $C$ (the second term on the
rhs of eq.\ (\ref{etfe}) corresponds to the charging energy of a classical
capacitor, corrected for the self-interaction of the excess charge 
\cite{yl1,yl2}).
We remark that regarding the system as a classical conductor, where the
excess charge accumulates on the outer surface, yields a value 
of $C=8.32$ $a.u.$ (that is the outer radius of the jellium shell).
Naturally, the ETF calculated value for $C$ is somewhat larger because of the
quantal spill-out of the electronic charge density.
Indeed, from the slope of $\widetilde{g}(z)$ we determine\footnote{ 
Due to the changing spill-out with excess charge $z$,
the capacitance should be written as $C+\delta(z)$. For our
purposes here the small correction $\delta(z)$ can be neglected.}
$C = 8.84$ $a.u$.

A similar plot of the shell-corrected and icosahedrally modified energy
differences 
$\Delta E^{\text{ico}}_{\text{sh}}(z)
=E^{\text{ico}}_{\text{sh}}(z)-E^{\text{ico}}_{\text{sh}}(0)$
is shown in Fig.\ 4b (in the range $-2 \leq z \leq 4$, filled circles).
The function $g^{\text{ico}}_{\text{sh}}(z)$,
defined as in eq.\ (\ref{gsm}) but with the shell-corrected quantities
($\Delta E^{\text{ico}}_{\text{sh}}(z)$ and $A^{\text{ico}}_1$),
is included in the inset to Fig.\ 4a (filled circles).
The shift discernible between  $g^{\text{ico}}_{\text{sh}}(-1)$
and $g^{\text{ico}}_{\text{sh}}(1)$ is approximately 1.7 $eV$, and
originates from the difference of shell effects on the IPs and EAs (see
Table I). The effect of shell-closures for $z=-10$ and $z=6$ 
(which was discussed earlier 
in the context of Fig.\ 3b for higher IPs and EAs) is not discernible
in the case of $g^{\text{ico}}_{\text{sh}}(z)$ due to the scale of the inset. 
The segments of the curve $g^{\text{ico}}_{\text{sh}}(z)$
in the inset of Fig.\ 4a, corresponding to positively ($z < 0$) and
negatively ($z > 0$) charged states, are again well approximated by straight 
lines, whose slope is close to that found for $\widetilde{g}(z)$.
Consequently, we may approximate the charging energy,
including shell-effects, as follows,
\begin{equation}
E^{\text{ico}}_{\text{sh}}(x) = E^{\text{ico}}_{\text{sh}}(0)
+ \frac{x(x-1)e^2}{2C} - A^{\text{ico}}_1 x~,
\label{eshneg}
\end{equation}
for {\it negatively\/} charged states, and 
\begin{equation}
E^{\text{ico}}_{\text{sh}}(x) = E^{\text{ico}}_{\text{sh}}(0)
+ \frac{x(x-1)e^2}{2C} + I^{\text{ico}}_1 x~,
\label{eshpos}
\end{equation}
for {\it positively\/} charged states. Note that without shell-corrections
(i.e., ETF) $\widetilde{I}_1-\widetilde{A}_1 = e^2/C = 27.2/8.84\; eV 
\approx 3.1 \; eV $, because of the symmetry of eq.\ (\ref{etfe}) with respect
to $z$, while the shell-corrected quantities are related as 
$I^{\text{ico}}_1-A^{\text{ico}}_1 \approx e^2/C + \Delta_{sh}$,
where the shell correction is $\Delta_{sh} \approx 1.55\; eV$
(from TABLE I, $I^{\text{ico}}_1-A^{\text{ico}}_1 \approx 4.65 \; eV$).

Expression (\ref{eshneg}) for the negatively charged states can be rearranged
as follows (energies in units of $eV$),
\begin{equation}
E^{\text{ico}}_{\text{sh}}(x)-E^{\text{ico}}_{\text{sh}}(0)=
-2.99 + 1.54 (x-1.39)^2~,
\label{eshsq}
\end{equation}
in close agreement with the all-electron LDA result of 
Ref.\ \cite{pede}. 

Equations (\ref{eshneg}) and (\ref{eshpos}) can be used to provide simple
analytical approximations for the higher IPs and EAs (see the definition
in eqs.\ (\ref{ips}) and (\ref{eas})). Explicitly written,
$A^{\text{ico}}_x \equiv E^{\text{ico}}_{\text{sh}}(x-1) -
E^{\text{ico}}_{\text{sh}}(x) = A^{\text{ico}}_1  - (x-1)e^2/C$ and
$I^{\text{ico}}_x = I^{\text{ico}}_1  + (x-1)e^2/C$.
Such expressions have been used previously \cite{wang} with an assumed value 
for $C \approx 6.7\; a.u.$ (i.e., the radius of the C$_{60}$ molecule, as
determined by the distance of carbon nuclei from the center of the 
molecule), which is appreciably smaller than the value obtained by us
($C=8.84 \; a.u.,$ see above) via a microscopic calculation.
Consequently, using the above expression with our calculated value
for $A^{\text{ico}}_1 = 2.75 \; eV$ (see TABLE I), we obtain 
an approximate value of $A^{\text{ico}}_2 = -0.35 \; eV$ (compared to the 
microscopically calculated value of $-0.09\; eV$
given in TABLE I, and $-0.11\; eV$ obtained by Ref.\ 
\cite{pede}) --- indicating metastability of C$_{60}^{2-}$ ---
while employing an experimental value for 
$A^{\text{ico}}_1 = 2.74\; eV$, a value of $A^{\text{ico}}_2 = 0.68\; eV$
was calculated in Ref.\ \cite{wang}. 

Concerning the cations, our expression (\ref{eshpos}) with a calculated
$I^{\text{ico}}_1 = 7.40\; eV$ (see TABLE I) and $C=8.84 \; a.u.$ 
yields approximate values 18.5 $eV$ and 31.5 $eV$ for the appearance energies 
of C$_{60}^{2+}$ and C$_{60}^{3+}$ (compared to the microscopic
calculated values of 17.71 $eV$ and 30.99 $eV$, respectively,
extracted from TABLE I, and 18.6 $eV$ for the former obtained in Ref.\
\cite{rose}). Employing an experimental value for 
$I^{\text{ico}}_1 = 7.54 \; eV$, corresponding values of 19.20 $eV$ and
34.96 $eV$ were calculated in Ref.\ \cite{wang}. As discussed in Ref.\
\cite{baba}, these last values are rather high, and the origin of the
discrepancy may be traced to the small value of the capacitance which was
used in obtaining these estimates in Ref.\ \cite{wang}.

A negative value of the second affinity indicates that C$_{60}^{2-}$ is
unstable against electron autodetachment. In this context, we note that 
the doubly negatively charged molecule C$_{60}^{2-}$ has been observed in 
the gas phase and is believed to be a long-lived metastable species
\cite {hett,limb}. Indeed, as we discuss
in the next section, the small LDA values of $A^{\text{ico}}_2$ found by us
and by Ref.\ \cite{pede} yield lifetimes which are much longer than those
estimated by a pseudopotential-like Hartree-Fock model
calculation \cite{hett}, where a value of $\sim$ 1 $\mu s$ was estimated.

\subsection{Lifetimes of metastable anions, C$_{60}^{x-}$}

The second and higher electron affinities of C$_{60}$ were found
to be negative, which implies that the anions C$_{60}^{x-}$ with
$x \geq 2 $ are not stable species, and can lower their
energy by emitting an electron. However, unless the number of excess
electrons is large enough, the emission of an excess electron involves
tunneling  through a barrier. Consequently, the moderately
charged anionic fullerenes can be described as metastable species 
possessing a decay lifetime.

To calculate the lifetime for electron autodetachmant, it is necessary
to determine the proper potential that the emitted electron sees as
it leaves the molecule. The process is analogous to alpha-particle
radioactivity of atomic nuclei. The emitted electron will have a
final kinetic energy equal to the negative of the corresponding higher
EA. We estimate the lifetime of the decay process by using the WKB method,
in the spirit of the theory of alpha-particle radioactivity, which has
established that the main factor in estimating lifetimes is
the relation of the kinetic energy of the emitted particle to the 
Coulombic tail, and not the details of the many-body problem in the 
immediate vicinity of the parent nucleus.

Essential in
this approach is the determination of an appropriate single-particle potential
that describes the transmission barrier. It is well known that the LDA
potential posseses the wrong tail, since it allows for the electron to
spuriously interact with itself. A more appropriate potential would be one
produced by the Self-Interaction Correction (SIC) method of Ref. \cite{perd2}.
This potential has the correct Coulombic tail, but in the case of
the fullerenes presents another drawback, namely Koopman' s theorem is not
satisfied to an extent adequate for calculating lifetimes.\footnote{
For certain systems, such as for example sodium clusters, an 
orbitally-averaged-like SIC treatment yielded 
highest-occupied-molecular-orbital (HOMO) energies for anions in 
adequate agreement with the calculated electron affinities (see, Refs.\
\cite{yl1,yl2}).} 
In this context, we note that Koopman' s theorem is known to be poorly
satisfied for the case of fullerenes even in Hartree-Fock calculations 
\cite{cios}. Therefore, the HOMO corresponding to the emitted electron,
calculated as described above, cannot be used in the WKB tunneling
calculation.

Since the final energy of the ejected electron equals the negative of
the value of the electron
affinity, we seek a potential that, together with the icosahedral 
perturbation, yields a HOMO level in C$_{60}^{x-}$ with energy 
$-A_x^{\text{ico}}$.
We construct this potential through a self-interaction correction to the LDA
potential as follows,
\begin{equation}
V_{\text{WKB}} = V_{\text{LDA}}[\widetilde{\rho}] - 
V_{H}[\frac{\widetilde{\rho}}{N_e}] -
V_{\text{xc}}[\xi\frac{\widetilde{ \rho}}{N_e}]~,
\label{sicxi}
\end{equation}
where the parameter $\xi$ is adjusted so that the HOMO level of C$_{60}^{x-}$
equals $-A_x^{\text{ico}}$.
In the above expression, the second term on the rhs is an average 
self-interaction Hartree correction which ensures a proper long-range
behavior of the potential (i.e., correct Coulomb tail), and
the third term is a correction to the short-range exchange-correlation.

\begin{figure}[t]
\centering\includegraphics[width=6.5cm]{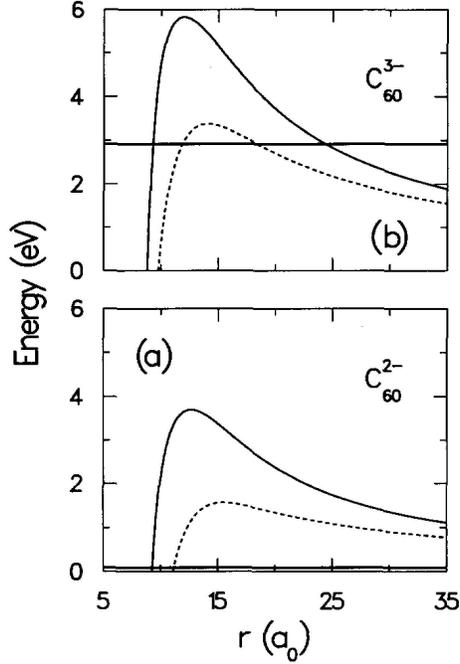}
\caption{
WKB effective barriers used to estimate lifetimes for
C$_{60}^{2-}$ (a) and C$_{60}^{3-}$ (b). Dashed lines correspond to barriers 
due solely to Coulombic repulsion and solid lines to total barriers after adding
the centrifugal components. The thick horizontal solid lines
correspond to the negative of the associated electron
affinities $A^{\text{ico}}_2$ (a) and  $A^{\text{ico}}_3$ (b).
In the case of C$_{60}^{2-}$ [panel (a)],
the horizontal solid line at $-A^{\text{ico}}_2=0.09\; eV$ crosses the total
barrier at an inside point $R_1 = 9.3 \; a.u.$ and again at a distance  
very far from the center of the fullerene molecule, namely at an 
outer point $R_2 = -e^2/A^{\text{ico}}_2 = 27.2/0.09\; a.u. = 302.2\; a.u.$
This large value of $R_2$, combined with the large centrifugal barrier,
yields a macroscopic lifetime for the metastable C$_{60}^{2-}$
(see text for details).
}
\end{figure}

For the cases of C$_{60}^{2-}$ and C$_{60}^{3-}$ such potentials are
plotted in Fig.\ 5. We observe that they have the correct
Coulombic tail, namely a tail corresponding to one electron for
C$_{60}^{2-}$ and to two electrons for C$_{60}^{3-}$.
The actual barrier, however, through which the electron tunnels
is the sum of the Coulombic barrier plus the contribution of
the centrifugal barrier. As seen from Fig.\ 5, 
the latter is significant, since the HOMO in the fullerenes 
possesses a rather high angular momentum ($l=5$), while being
confined in a small volume.

Using the WKB approximation \cite{baz},
we estimate for C$_{60}^{2-}$ a 
macroscopic half-life of $\sim 4 \times 10^{7} $ $years$, 
while for C$_{60}^{3-}$
we estimate a very short half-life of $2.4 \times 10^{-12}\; s$.
Both these estimates are in correspondence with observations. Indeed,
C$_{60}^{3-}$ has not been observed as a free molecule, while
the free C$_{60}^{2-}$ has been observed to be long lived \cite{hett,limb} 
and was detected even 5 $min$ after its production through
laser vaporization \cite{limb}.

We note that the WKB lifetimes calculated for tunneling 
through Coulombic barriers are very sensitive 
to the final energy of the emitted particle and can vary by many orders of 
magnitude as a result of small changes in this energy, a feature well known
from the alpha radioctivity of nuclei \cite{baz}.
Since the second electron affinity of C$_{60}$ is small, effects due
to geometrical relaxation and spin polarization
can influence its value and, consequently, the estimated lifetime. 
Nevertheless, as shown in Ref.\ \cite{pede}, inclusion of such corrections 
yields again a negative second affinity,
\footnote{The sign conventions in Ref.\ \cite{pede} are the opposite of
ours.}
but of somewhat smaller magnitude, resulting in an even longer lifetime. 
Furthermore, as discussed in Ref.\ \cite{coul}, the stabilization effect
of the Jahn-Teller relaxation for the singly-charged ion is only of the
order of 0.03 -- 0.05 $eV$. Since this effect is expected to be largest
for singly-charged species, C$_{60}^{2-}$ is not expected to be 
influenced by it \cite{pede}.

On the other hand, generalized
exchange-correlation functionals with gradient corrections yield slightly 
larger values for the second electron affinity. For example, using
exchange-correlation gradient
corrections, Ref.\ \cite{pede} found $A^{\text{ico}}_2 = - 0.3 \; eV$,
which is higher (in absolute magnitude) than the value obtained without 
such corrections. This value of $-0.3 \; eV$ leads to a much 
smaller lifetime than the several million of years that correspond to the 
value of $-0.09 \; eV$ calculated by us. Indeed, using the barrier displayed in 
Fig.\ 5a, we estimate a lifetime for C$_{60}^{2-}$ of approx. 0.37 $s$,
when $A^{\text{ico}}_2 = - 0.3 \; eV$. 
We stress, however, that even this lower-limit value still corresponds to
macroscopic times and is 5 orders of magnitude larger than the estimate of
Ref.\ \cite{hett}, which found a lifetime of 1 $\mu s$ for 
$A^{\text{ico}}_2 = - 0.3 \; eV$, since it omitted the large centrifugal 
barrier. Indeed, when we omit the centrifugal barrier, we find a lifetime
estimate of 1.4 $\mu s$, when $A^{\text{ico}}_2 = - 0.3 \; eV$.

\section{Summary}

We developed a new LDA method for investigation of complex carbon clusters, 
and, as an illustration, applied it to multiply charged fullerenes 
C$_{60}^{x \pm}$. The main elements of this method are: (i) Use of 
a {\it stabilized\/} jellium (structureless pseudopotential) 
approximation \cite{perd} for the ionic background, instead of the standard
jellium approximation;
(ii) Use of a recently introduced, shell-correction method \cite{yl1,yl2};
(iii) Inclusion of the effect of the icosahedral symmetry via a perturbative
treatment.

For the neutral C$_{60}$ molecule, the results obtained by our method are
in good agreement with experimental observations, as well as with previous
self-consistent electronic structure calculations \cite{trou,rose,ye}.

For the multiply charged fullerenes, charging energies were
microscopically calculated for up to $x=12$ excess charges. 
A semiclassical interpretation of these results was developed, which viewed 
the fullerenes as Coulomb islands \cite{kast1,kast2} possessing
a classical capacitance. A value of 8.84 $a.u.$ was found for this
capacitance, in contrast to the much smaller value of 6.7 $a.u.$ inferred
previously from the particle-on-a-sphere model \cite{wang}.
The calculated values for the first ionization 
potential and the first electron affinity agree well with the experimental 
ones. For the second and third ionization potentials, there exist substantial 
discrepancies in the experimental measurements \cite{baba}. 
Our calculations, reflecting the more accurate value of the capacitance of
C$_{60}$, support the results from charge transfer bracketing 
experiments \cite{elva} and from direct ionization experiments by electron 
impact \cite{baba}.

Employing the analogies with the case of alpha-particle radioactivity in nuclei,
we found that the doubly charged negative ion is a very long-lived metastable 
species, in agreement with observations \cite{hett,limb}.
It decays through spontaneous electron emission, and 
its macroscopic lifetime is the result of a superposition of a Coulombic
barrier and a large centrifugal barrier, through which the emitted electron
tunnels.

\section*{Acknowlegdments}

This research is supported by the U.S. Department of Energy, Grant No.
AG05-86ER45234.

\end{document}